# THz pulses from optically excited Fe, Pt and Ta based Spintronic Heterostructures

Sandeep Kumar, Anand Nivedan, Arvind Singh, and Sunil Kumar*

*Femtosecond Spectroscopy and Nonlinear Photonics Laboratory,*
*Department of Physics, Indian Institute of Technology Delhi, New Delhi 110016, India*
*\*Email: kumarsunil@physics.iitd.ac.in*

Spintronic heterostructures are considered to be the new generation THz sources for their capability in producing high power and broadband THz radiation. Here, we provide a brief review on the state-of-the-art in this field. The optically excited bi- and tri-layer combinations of ferromagnetic and nonmagnetic thin films have become increasingly popular. Towards optimizing the THz conversion efficiency and broadband gapless spectrum from these THz emitters, various control parameters need to be taken into consideration. The inverse spin Hall effect in the heavy metal layer of the heterostructure is primarily responsible for the generation of THz pulses. A few new results on iron, platinum and tantalum based heterostructures have also been reported here. It is observed that the Ta(2nm)/Fe(2nm)/Pt(2nm) tri-layer heterostructure generates ~40(250)% stronger THz signal as compared to the counterpart Fe(2nm)/Pt(2nm) (Fe(3nm)/Ta(2nm)) bi-layer heterostructure.

## 1. Introduction

From electronic to optoelectronic sources, the generation of terahertz (THz) radiation covering ~0.1 to 20 THz frequency band in the electromagnetic spectrum, is realized these days by a number of ways [1-6]. These include thermal sources [7], vacuum electronic sources [8], THz lasers [9], and the optically pumped sources including the air plasma and nonlinear optical crystals [10, 11]. From the technique that is based on photomixing [12, 13], a tunable and narrowband continuous wave THz radiation can be generated. These types of sources are limited in terms of the bandwidth and the power they can provide. Electron accelerator-based sources can provide tunable bandwidth and very high peak-electric field of a few MV/cm at a repetition rate of a few MHz [14]. However, table-top THz sources, which can provide high field-strength along with the large bandwidth, are always on the top priority. Due to immense developments in the field of ultrafast lasers in last two decades, the ultrafast laser-driven table-top THz sources have become widely popular. THz sources based on the optical rectification in various optical crystals can also provide high peak-field at low pulse repetition rates [15]. High peak-field THz radiation at frequencies above 5 THz can also be provided by the optical parametric amplifiers [16]. Air plasma photonics-based sources can provide nearly gapless broadband THz pulses. [17]

Most of the THz sources as discussed above are based on ultrafast laser excited semiconductors and insulators where charge degree of freedom of an electron is being utilized. The tremendous progress in the field of spin-based electronics [18, 19] and unique virtues of spin degree of freedom of the electron, i.e., less heat dissipation and high operation speed [20], motivated researchers to take advantage of this in the ultrafast THz generation. A paradigm shift took place in the way THz pulses are generated from femtosecond (fs) laser excited materials due to the ultrafast inverse spin Hall effect [21-25]. Spintronic heterostructures of ferromagnetic (FM) and nonmagnetic (NM) metallic thin films are becoming widely popular as the new generation THz emitters for the large scope they provide in achieving high power and gapless ultrabroad band THz emission [22, 26]. In most of those studies on the spintronic THz emitters (STEs), nano-Joules of fs laser pulse energies at the visible/near-infrared (NIR) have been utilized to optically pump the heterostructures. Under the optimized conditions of the excitation pulse duration, film thicknesses and their appropriate combinations can produce THz pulses containing a broad spectrum and the THz power generation efficiencies quite comparable to or even larger than those from the popular semiconductor-based photoconductive antennas [11]. The STEs also exhibit unique properties such as the short electron lifetime [27] and featureless THz refractive index [28], which help in extremely fast transient photocurrent and gapless broadband THz emission from them. Moreover, the STEs are compact, scalable, and low-cost alternatives.

In this article, we have reviewed the latest state-of-art in the THz radiation generation from spintronic heterostructures that are optically pumped by ultrafast visible and NIR pulses along with a few new results on bi- and tri-layer spintronic heterostructures. In Section 2, we first discuss the underlying mechanism for the generation of THz pulses from optically excited FM/NM thin film heterostructures. Various material parameters and the experimental conditions with the optical excitation, which, determine the characteristics of the emitted THz radiation, are discussed in Section 3. Finally, in Section 4, we have presented new results on iron (Fe), platinum (Pt) and tantalum (Ta) based bi- and tri-layer STEs grown by using ultra high vacuum radio frequency (RF) magnetron sputtering. They are optically pumped by sub-100fs pulses from a low repetition rate chirped pulse amplifier operating at 800nm. A comparative study has been made for the THz generation efficiencies of Ta(2nm)/Fe(2nm)/Pt(2nm), Fe(2nm)/Pt(2nm) and Fe(3nm)/Ta(2nm) STEs grown on different substrates by keeping the experimental conditions with the optical excitation and electro-optic detection the same. In conjunction with the time-domain THz spectroscopy, the STEs provide an opportunity to determine various spintronic parameters of the underlying ferromagnetic and heavy metal layers used in the heterostructures in an all-optical manner [29]. They can also be exploited during the experiments for studying the ultrafast magnetization dynamics in the THz frequency range. [30-32]

## 2. Spintronic THz Emitters: The Principle

Beaurepaire *et al.,* [33] first observed THz generation from the nickel (Ni) thin films due to the ultrafast demagnetization under the excitation of femtosecond pulses. At the same time, the emission was also observed from metallic structures [34] (i.e., Ag and Au) and other FM thin film samples [35]. However, the signal strength emitted from alone, either ferromagnetic or metallic, was very weak. This is due to the absence of the spin sink. Because the THz generation through ISHE mechanism require both spin source (FM material) and efficient sink (heavy metal) [25]. By considering the above fact, the simplest structure of the spintronic based THz emitter is a bi-layer heterostructure consists a ferromagnetic and a heavy metal thin film layer and demonstrated in the number of works [21-23, 25]. As shown in the material schematic in Figs. 1(b) and 1(c), the basic principle behind the THz generation process from such bi-layer spintronics heterostructure is, the optical excitation in such FM/NM heterostructure by NIR femtosecond pulses, promotes electrons from below the Fermi level to above it in the FM layer and creates a nonequilibrium electron distribution [21]. Due to the difference in the density, band velocity, and lifetime of the spin-up (majority) and spin-down (minority) electrons in the FM layer [36], the equilibrium is achieved through launching an spin polarized current ($J_s$) in the super-diffusive manner [37]. The generated spin polarized current is then pumped from the FM layer into NM layer and due to strong intrinsic spin-orbit coupling in the NM layer (heavy metal), the spin-up and spin-down electrons are deflected oppositely by an amount of spin Hall angle to produce a net charge current density ($J_c$) by the virtue of inverse spin Hall effect (ISHE) [19]. This can be expressed through a simple relation [21, 25],

$$\vec{J_c} = \gamma \, (\vec{J_s} \times \hat{m}) \qquad (1)$$

Where, $\hat{m} = \vec{M}/|\vec{M}|$ is the initial magnetization direction, determined by the saturated magnetization of the FM layer and γ is the spin Hall angle. The spin to charge current conversion efficiency is mainly determined by the magnitude of this spin Hall angle, as depicted in Figs. 1(b) and 1(c).

The studies are not limited only to the bi-layers, but it is further extended through stack engineering to the tri-layer NM/FM/NM heterostructures concept. Seifert *el al.* [22], have shown that the tri-layer heterostructures are better than their bi-layers counterpart as it utilize the backward flowing spin current too (see Fig.1(c)), and their efficiently conversion to the THz radiation through ISHE in similar fashion as discussed in bi-layer. In their study, they have reported that the deposited STE tri-layer W(2nm)/CoFeB(1.8nm)/Pt(2nm) on glass substrate under the excitation of the Ti:Sapphire laser oscillator pulses (800nm, 25nJ energy, 80 MHz, 10fs duration), provide broadband THz spectrum exceeding from 1 to 30 THz along with the comparable field strength to that of ZnTe THz emitter. In a subsequent study [26], they have also reported the similar tri-layer combination of different optimized thickness (W(1.8nm)/CoFeB(2nm)/Pt(1.8nm)) on a large area glass substrate of diameter 7.5cm. Now, under the excitation with high energy amplified femtosecond NIR pulses (5.5mJ, 1kHz, 800nm centered, duration 10fs), generates a very high peak THz electric field of strength ~300 KV/cm and spanned the spectral bandwidth from 0.1 to 10 THz. The only change in the excitation can lead to weak to strong THz regime and suggests many more scope is available to achieve both the strong field and gapless emission of THz from this scheme.

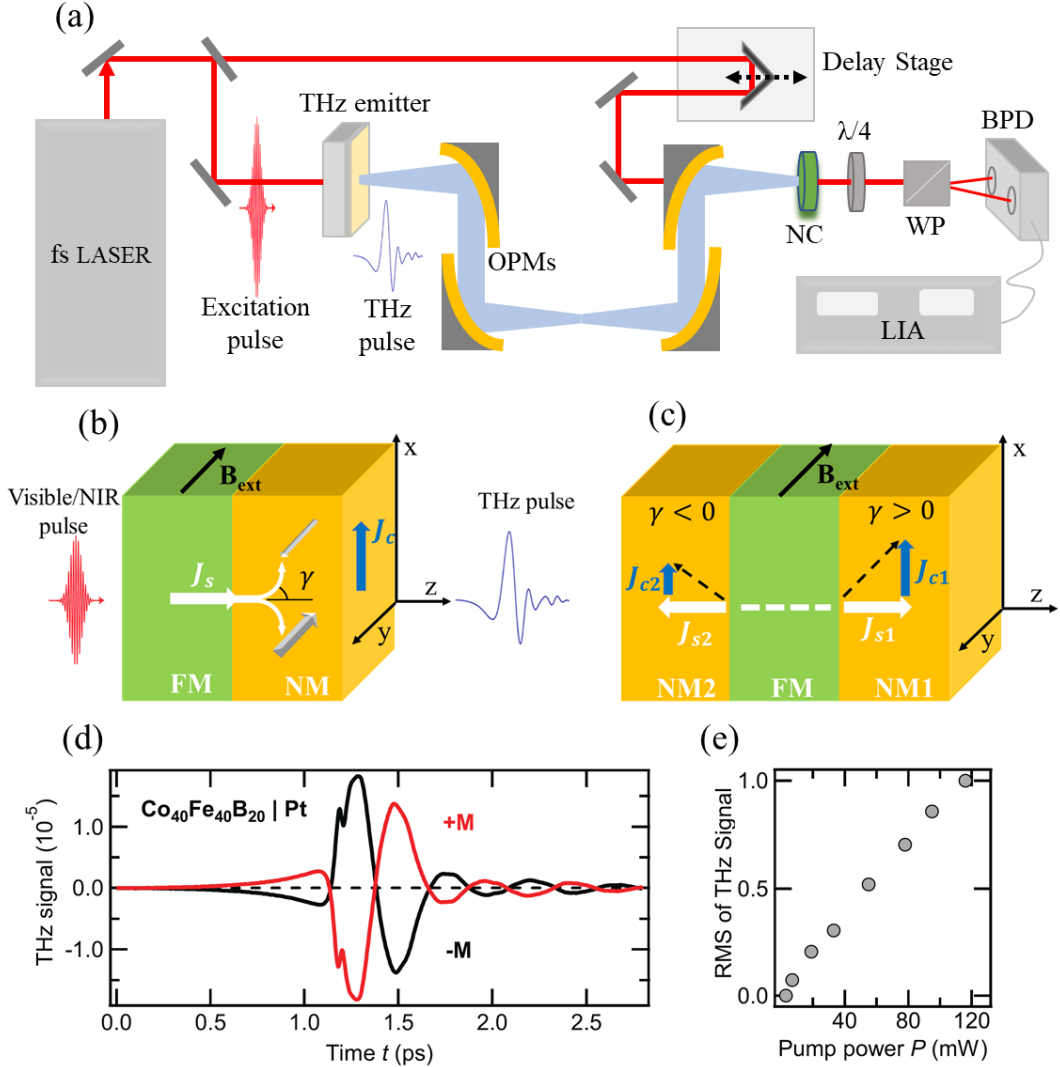

Fig.1. (a) Schematic of the experimental time-domain THz spectroscopy setup used for the generation and detection of the THz pulses from spintronic emitters. THz emission mechanisms from the (b) FM/NM bi-layer, and (c) NM2/FM/NM1 tri-layer spintronics heterostructure. (d) A typical time-domain THz electic field signal obtained from a CoFeB/Pt bi-layer structure for two opposite magnetization directions [29]. (e) THz signal magnitude dependence on the incident excitation pump power [29]. OPMs: off-axis parabolic mirrors, λ/4: quarter-wave plate, NC: nonlinear optical crystal, WP: Wollaston prism, BPD: balanced photodiode, LIA: lock-in amplifier, $B_{ext}$: external magnetic field, $J_s$: spin current, $J_c$: charge current, γ: spin Hall angle, FM: ferromagnetic, NM: nonmagnetic. Thick and thin arrows in the NM layer represent the oppositely moving majority and minority spin electrons, respectively, to produce transient charge current by virtue of inverse spin Hall effect in the NM layer.

## 3. Effect of Various Tunable Parameters

As discussed in the previous section, the use of mJ energy pulses instead of nJ laser pulses for the excitation of the similar spintronic heterostructure, can provide the tunability in the emitted THz electric field strength and bandwidth. Although, the low repetition rate pulses can generate high field THz radiation, but one has to compromise with the signal to noise (s/n) ratio they provide. Also, the optical to THz conversion efficiency of the spintronic emitters mainly depends on the material parameters. Similarly, there are various other tunable material and related spintronic parameters also, which controls the magnitude and bandwidth of the generated THz radiation [22, 29]. Therefore, proper optimization of these parameters is crucial to know and in order to improve the spintronics based THz device performance. Once full control is achieved on these parameters, an optimization in the THz generation efficiency for power and bandwidth is required through optimal conditions on the femtosecond pulse excitation and the respective optical setup by maximizing collection and detection of the THz radiation. Here, in the following subsections, these are discussed in detail.

### 3.1. Material Selection and Thicknesses

In very first optimization process, it is important to select a suitable material and their appropriate thickness combination. Here, in the case of spintronics THz emitters, number of ferromagnetic (Fe, Co, CoFeB, Ni, etc.) and nonmagnetic (Pt, Ta, W, Au, Pd, Ru, Ir, etc.) materials have been reported for the optically excited bi- and tri-layer heterostructure combinations [22, 38]. The appropriate combination of these FM and NM materials is then characterized by the generated THz signal efficiency from them. Figure 2 shows the variation in the THz signal amplitude under the replacement of the various NM layer while keeping constant the magnetic one. The results clearly indicate that the THz emission is highly sensitive to the selection of the proper NM material. However, it is observed that the replacement of the FM layer will not affect the THz signal amplitude much. The NM materials are mainly characterized according to their spintronics parameters and one very important of them is the spin-Hall angle [18, 39]. The spin hall angle of some NM materials is negative (W, Ta, etc.) and some of them have positive (Pt, Ru, Au, etc.) value. As the emitted THz signal is directly proportional to the spin Hall angle (Eq.1), so it is preferred to use the NM materials which possess large spin Hall angle as used by Seifert *et al.,* [22] in the optimized tri-layer heterostructure.

Along with the material selection, the layer thickness is another important parameter which needs to be handled very carefully in order to achieve maximum THz signal. The peak THz signal amplitude with respect to the varied thickness of FM and NM material while keeping fix the corresponding NM and FM layer thickness, respectively, have been shown in the previous reports [22, 23, 25]. The variation in both follows similar type of trend as it first increases to achieve a maximum for a particular optimized thickness and after that it starts to decrease. In the FM(varying)/NM(fixed) bi-layer heterostructures, the laser induced spin diffusion phenomenon boost the signal and then de-escalated by the THz absorption in thicker films. Similarly, due to the limited spin diffusion in the NM layer with the increased thickness in FM(fixed)/NM(varying) bi-layer configuration results in the similar kind of trend. To quantify the thickness dependance in STEs, in fact, the overall thickness of the spintronic heterostructure along with the generation and diffusion of hot carries in FM, spin accumulation in NM and the THz absorption in FM, has to be considered while optimizing the optical to THz conversion efficiencies of the STEs. The initial increment in the THz signal amplitude with the varied thickness of is due to increased efficiency of the generated spin current in FM layer and the consequent conversion into THz radiation. Whereas, after a certain thickness combination, the THz attenuation due to the increased stack thickness overcome this generation efficiency hence, decreased signal is observed for higher thickness. However, there are also many microscopic origins such as spin relaxation length, interface quality, transmission properties, magnetization, etc. [29], responsible for such type of variation observed in THz signal with FM/NM layer thicknesses.

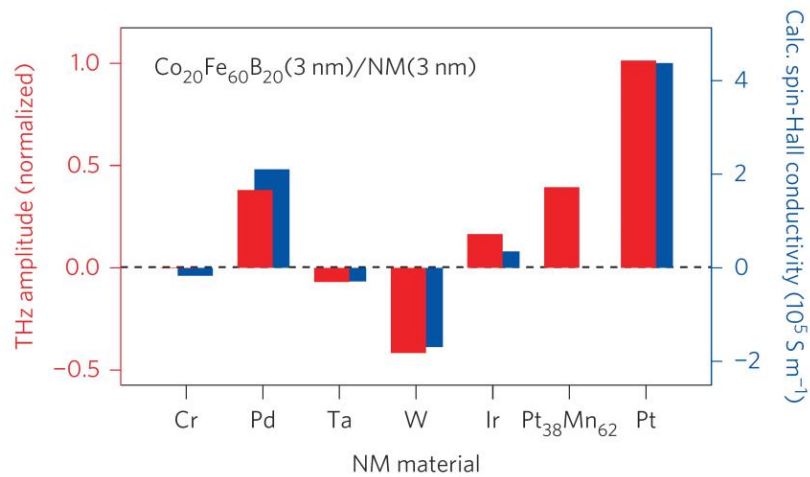

Fig.2. Normalized peak THz amplitude from a CoFeB(3nm)/NM(3nm) bi-layer heterostructure having different NM material layers (red bars). Also shown is the ab-initio calculated values of the spin Hall conductivity (blue bars). [22]

### 3.2. Optical Excitation – Polarization and Wavelength

The dependency of the emitted THz signal on the incident laser pulse polarization is basically defined by the type of involved mechanism which generates the THz radiation. If the THz is generated from FM/NM heterostructure by the ISHE process then the emitted THz transient is independent to the incident laser helicity [21, 25], can also be seen in Fig.3. Whereas, the THz emission due to the broken inversion symmetry results to the helicity dependent signal [40].

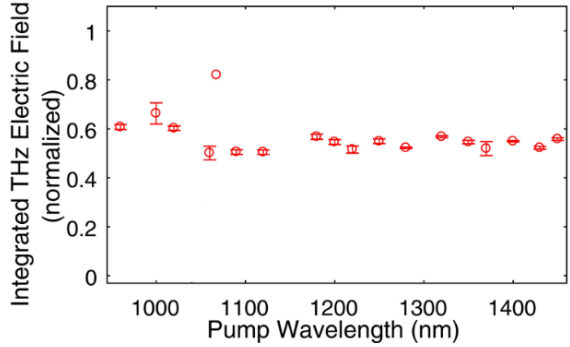

Fig.3. Normalized THz electric field from a tri-layer heterostructure (W/CoFeB/Pt) obtained by ultrafast optical excitation at different wavelengths from 900nm to 1500nm. [41]

The impact of varying excitation wavelength have been studied on spintronic bi- and tri-layer heterostructures [41, 42]. Most of them have shown that the THz signal is independent of the excitation wavelength and have linear dependency on incident power for a fixed wavelength. As shown in Fig.3 , Herapath *et al.* [41], investigated the effect of various pump wavelength in the infrared region ranging from 900 to 1550nm and found that the emission of THz signal is independent (nearly constant) of this incident wavelength range. Whereas, Adam *et al.* [43], have shown the contradictory results for the excitation wavelength of 400nm. A 3-fold THz signal enhancement is observed when the tri-layer heterostructure is excited with 400nm as compared to 800nm pulses. This kind of behavior is attributed to large concentration of hot electrons from the high energy blue (400nm) photons which results to the enhanced spin current efficiency.

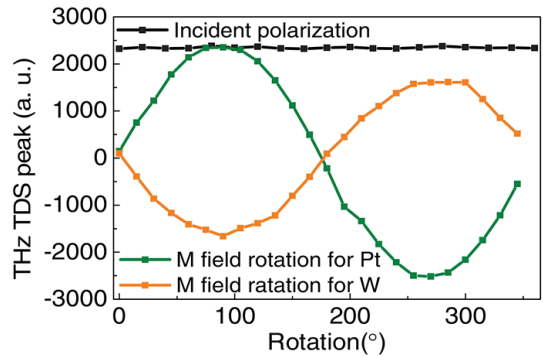

Fig.4. Peak THz electric field versus the polarization direction of the incident optical excitation pulse (black curve for Co(4nm)/Pt(4nm) STE on the $SiO_2$ substrate), and the direction of the in-plane external magnetic field (green curve for Co(4nm)/Pt(4nm) and olive curve for Co(4nm)/W(4nm), both on the $SiO_2$ substrates). [25]

### 3.3. Polarization of Emitted THz radiation

Another aspect of the STE is that the emission THz transient from these spintronics heterostructure is completely magnetic in origin and has direct relation with its magnetization direction as can be seen from Eq.1. As discussed with details in Section 2, the polarization of the generated THz radiation can be fully controlled by the applied external magnetic field and has been show in previous reports also [21, 24]. For instance, Wu *et al.* [25], have shown the variation of the THz time-domain signal with the rotation of the externally applied magnetic field follow a sinusoidal type behavior in the full $360^0$ rotation. The simple magnetic controlled THz polarization in such emitters eliminates the requirement of expensive THz polarizer. The sign reversal of the THz signal is also observed whether the sample orientation or the excitation direction is flipped. [23, 44]

**4. Bi- and Tri-layer Spintronic THz Emitters for High Pulse-Energies**

As pointed out earlier, the optical to THz conversion efficiency of the spintronic THz emitters primarily depends on the FM and NM material parameters for given experimental conditions with the excitation laser and detection. By increasing the excitation pulse energy, one can obtain THz pulses having energy increased in the same proportion. In this section, we have presented a few results on Fe, Ta and Pt based bi- and tri-layer spintronic THz emitters which are optically excited by amplified laser pulses from a low-repetition rate femtosecond laser at 800nm.

**4.1. Characterization of the Materials and Ultrafast Experiments**

Thin film samples of Fe/Pt, Fe/Ta and Ta/Fe/Pt heterostructures were fabricated on different substrates by using ultra-high vacuum RF magnetron sputtering. The base pressure was kept at ~$4\times10^{-8}$ Torr. Substrates of fused silica (quartz) and sapphire (0001) were used. In order to remove the contamination from the surfaces of the substrates, they were pretreated with acetone and isopropyl alcohol (IPA) in ultrasonic bath. The working pressure during the thin film deposition in the Argon environment was kept in the range of ~2-5 mTorr, depending upon the depositing material and to achieve precise control over surface and interface roughness. The growth rates for various deposited layers of the magnetic and nonmagnetic metallic materials were optimized for a particular power of the RF supply. The optimized growth rates were ~0.09 Å/sec, 0.44 Å/sec, and 0.76 Å/sec for Fe, Pt and Ta, respectively. Combinations of Fe, Pt and Ta were deposited in sequence to obtain the required stack of bi- and tri-layer heterostructures. For the initial structural characterization, X-ray diffraction (XRD) measurements were performed using the PANalytical X'Pert diffractometer with Cu-K$_\alpha$ source. The obtained XRD patterns (not shown) for the nonmagnetic layers used in the heterostructures, i.e., the Pt and Ta films, confirms that the growth takes place primarily in polycrystalline phase and α-phase, respectively. The phase confirmation results are also consistent with those in the previously reported literature on Pt [45] and Ta [46]. The same system was also used for surface and interface roughness measurement in the X-ray reflectivity (XRR) mode. Although, the individual films were deposited with the corresponding optimized growth rates, but the further confirmation of the thickness and roughness was obtained from fitting of the XRR plots (not shown here) using the recursive theory of Parrat [47]. The magnetic properties of the samples were studied using vibrating sample magnetometer (VSM) of a physical property measurement system (Quantum Design).

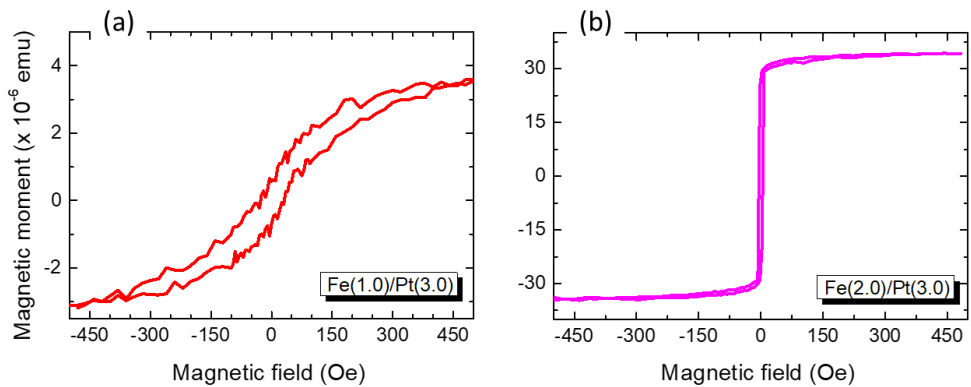

Fig.5. Hysteresis loops from magnetic measurements using vibrating sample magnetometer on Fe/Pt bi-layer spintronics heterostructures. Thickness of Pt film is kept fixed at 3.0nm and that of the Fe film in the heterostructure is changed from (a) 1.0nm to (b) 2.0nm.

Saturation magnetization and easy axis of the magnetization were determined from the magnetic hysteresis loop (M-H) measurements. For this, experiments on films having different thicknesses were performed. Fig. 5 presents results from such measurements on two heterostructures of Fe/Pt using in-plane external magnetic field varying from -500 Oe to +500 Oe. For Fe(1.0)/Pt(3.0) in Fig. 5(a), where numbers inside the small parentheses are film thicknesses in nm, a linear non-saturating magnetization can be seen. On the other hand, for Fe(2.0)/Pt(3.0) in Fig.

5(b), clear hysteresis loop with saturation magnetization is observed. These results indicate the presence of the in-plane easy axis component for the Fe film at a thickness between 1 and 2nm. The critical thickness for Fe has been reported to be ~1.5nm in the literature [48]. For Fe film thicknesses lesser than 1.5nm, there is a possibility of having a magnetization component that is normal to the surface [49]. However, we could not confirm this due to limitations in our experiments. It is worth mentioning that the out of plane magnetization component does not affect our results for the generation of THz pulses from the spintronics structures as discussed later in the paper. From the M-H measurements, we calculate value of the coercive field to be ~30 Oe. Therefore, in all our THz experiments, we have kept the in-plane external magnetic field appropriately, i.e., well above the saturation field value to ensure perfectly magnetized samples.

The THz experiments were carried out using a time-domain setup that is schematically shown in Fig. 1(a). The THz emitter in this case is the spintronics heterostructure as discussed above. A Ti:sapphire regenerative amplifier of 1kHz pulse repetition rate was used. The laser pulses have full-width at half maximum (FWHM) duration of ~35fs and their spectrum is centered at 800nm. The output of the laser is divided into 90/10 ratio using a beam splitter. The strong part was used to optically excite the spintronic heterostructure while the weak one was sent through a time-delay stage to an optical crystal for detecting the THz pulses by electro-optical sampling [50]. For all the THz results presented here, the excitation and sampling beam powers were kept constant throughout the experiments. Pump pulse energy and beam diameter at the sample point were ~350μJ and ~6mm, respectively. The sampling beam pulse energy was kept at <0.4μJ while the beam diameter was ~2mm. As discussed in Section 2, through conversion of spin current to charge current in the nonmagnetic heavy metal layer (either Pt or Ta in our case), THz pulses are emitted from the optically excited spintronic heterostructure. The generated THz pulses are then routed through a set of four parabolic mirrors (OPMs) and towards the electro-optic crystal for detection. The residual optical beam is separated from the THz beam using a high resistive silicon wafer immediately after the THz emitter. For proper detection of the THz pulses, a <110>-oriented ZnTe crystal, a quarter wave plate, a Wollaston prism, and a balance photo detector arrangement were used. The difference signal in the balanced photodiode is detected on a lock-in amplifier and it is directly proportional to the THz electric field. Therefore, comparative studies on the THz generation efficiencies among various THz emitters can be performed in our experiments keeping all the conditions with the optical excitation the same. All the experimental results reported here have been achieved under the same atmospheric conditions of the temperature (~25 °C) and humidity (~40%).

Bi-layer FM/NM and tri-layer NM2/FM/NM1 heterostructures were used as the THz emitters in our experiments. The THz generation process from a bi-layer and tri-layer spintronics heterostructure is already described in section 2 and depicted in Fig1. (b) and (c), respectively. The sample is magnetized along the -y-direction. The in-plane transient charge current burst ($J_c$) produces THz radiation, whose polarization can be controlled by the direction of magnetization of the sample [23] as discussed in Section 3.3. We confirmed that by reversing the direction of the external magnetic field, the polarity of the emitted THz time-domain signal, E(t), also gets inverted without any explicit changes in the temporal shape of the THz pulse and its peak value, is consistent with the results described in Section 3.3. Clearly, upon flipping the direction of the external field, the spin polarization direction gets reversed and so is the polarity of the transient charge current, $J_c$. The optical excitation of all the samples studied here, were done from the substrate side. The geometric details are: Substrate/Fe/Pt(or Ta) for the bi-layers and Substrate/Ta/Fe/Pt for the tri-layers.

## 4.2. THz Radiation from Fe/Pt, Fe/Ta and Ta/Fe/Pt: A Comparison Between Bi- and Tri-layer Spintronic Emitters

THz signals from the optimized Ta(2nm)/Fe(2nm)/Pt(2nm) tri-layer and counterpart Fe(2nm)/Pt(2nm) and Fe(3nm)/Ta(2nm) bi-layer FM/NM heterostructures are shown in Fig. 6(a), while, Fig. 6(b) presents the corresponding Fourier spectra. Here, quartz is the substrate. Due to strong THz absorption by moisture in air, there are dips in the THz spectra (Fig. 6(b)) at certain frequencies corresponding to the water molecular vibrations [51]. However, under constant environmental conditions, those do not hinder the main observations and conclusions drawn in our paper. In the absence of an appropriate thickness heavy metal (Pt or Ta) layer, the ferromagnetic (Fe) layer, capped with an extremely thin nonmagnetic and non-oxidizing material layer, can also contribute to the THz emission process. However, this process of THz generation from the Substrate/FM structure due to optically induced ultrafast demagnetization in the FM layer is extremely weak [35]. Having an appropriate thickness heavy metal layer on one or either side of the FM layer is the ideal structure for THz emission in practical applications. The optical to THz frequency conversion in such spintronics structures depends on the overall thickness of the FM/NM combinations [22, 25].

While comparing the THz signals from the Fe/Pt and Fe/Ta bi-layers under the same experimental conditions, it can be seen from Fig. 6(a) that not only the polarity of the signal gets reversed in the case of Fe/Ta but also the magnitude is few times smaller than that from Fe/Pt. The total thickness of the FM/NM bi-layer structures in both the cases is kept the same. Having the same spintronics material layer thicknesses and the substrate underneath, a direct comparison of the THz signals from the Fe/Ta and the Fe/Pt bi-layers, only in terms of the spintronics parameters, is quite reasonable. The negative spin Hall angle, $\gamma$ and differences in other spintronic parameters of the Ta layer in comparison to Pt [46, 52, 53], are responsible for the observed THz signals from the Fe/Pt and Fe/Ta bi-layers in Figs. 6(a) and 6(b). We may point out that the interface quality between the FM and NM layers also contributes in the overall THz conversion efficiency from such heterostructures [54].

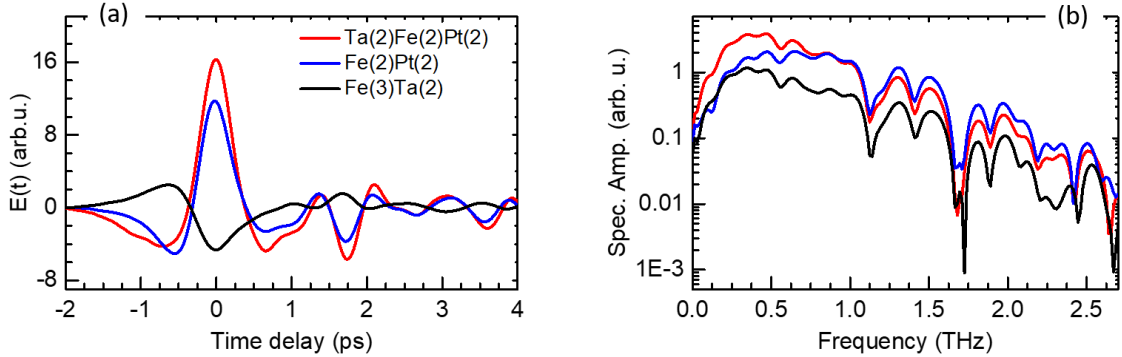

Fig.6. Comparison between THz generation efficiency of bi- and tri-layer spintronic heterostructures using optimized thickness combinations of the magnetic (Fe) and nonmagnetic layers, either Pt or Ta. (a) Time-domain THz signal, E(t) and (b) Fourier transformed field spectra.

Figure 6 also presents a comparison between the THz generation efficiencies from the bi-layer structure with another tri-layer structure, both having optimized thicknesses of the FM and the NM layers on quartz substrates. Ta/Fe/Pt heterostructure provides stronger THz signal as compared to the bi-layer counterparts, Fe/Pt or Fe/Ta, though the bandwidth in the spectrum appears quite similar. The reason for the stronger THz signal from the tri-layer than the bi-layer is the utilization of the backward flowing spin current in the second nonmagnetic layer as also discussed in Section 2. In the FM/NM bi-layer heterostructures, the ultrafast photoexcitation generated spin current propagates in all possible directions, and only half of the spin current from Fe layer is converted into charge current in the adjacent Pt layer on one side. On the other hand, in the tri-layer Ta/Fe/Pt heterostructure, the two interfaces work in tandem so as the THz radiation from Ta/Fe interface adds up coherently with that from the Fe/Pt interface. Therefore, this type of geometrical engineering of FM and NM layers provides ~40% and ~250% enhanced THz signal from the Ta/Fe/Pt as compared to the Fe/Pt and the Fe/Ta, respectively. Note that the overall thickness difference in both the structures is around 50%. Similar enhancement in the THz signal from W/CoFeB/Pt tri-layer as compared to the CoFeB/Pt bi-layer structure was reported in the literature where high repetition rate and low pulse energy excitation pulses were used [22]. The results clearly indicate that the addition of the Ta layer in the tri-layer structure enhances the THz generation efficiency due to the utilization of backflow spin current as well and their conversion into THz charge current through ISHE. The polarity of the THz signal from the Ta(2)/Fe(2)/Pt(2) is similar to the Fe(2)/Pt(2).

**4.3 Role of the Substrate**

The choice of appropriate substrates is a crucial factor in the optimization of the spintronics THz emitters as it affects the temporal and frequency domain shapes and overall magnitude of the THz signal [23, 25]. In Fig. 7, we have compared the THz signals generated from the Ta(2nm)/Fe(1.5nm)/Pt(2nm) tri-layer heterostructures deposited on two different substrates, i.e., 1mm thick quartz and sapphire. The time-domain signals obtained from both the samples have almost same peak value but a slight difference in their temporal shapes. It is observed that the signal from the tri-layer sample on the sapphire substrate has slightly narrower temporal shape than the sample on the quartz substrate. This may be due to the difference in the thin film growth template in the two cases and their absorption characteristics [23]. We must point out that the optical and THz transmissions of both the substrates were nearly the same.

The spectral bandwidth of the generated THz radiation from all the bi- and tri-layer spintronic heterostructures appears limited up to a frequency of ~3 THz. Though, we have used sub-100fs NIR pulses for optical excitation in our experiments, we believe that the limited bandwidth of the generated THz radiation mainly arises from the 1mm thick ZnTe crystal that was used for the detection of the THz pulses by electro-optic sampling. More experiments with thinner crystals are being carried out to clarify this further. Nevertheless, our findings suggest that the spintronics THz emitters work even with the amplified fs laser pulses and there is substantial scope for the improvement in the THz generation from them so as to provide powerful and broadband THz emitters for futuristic needs in THz science and technology.

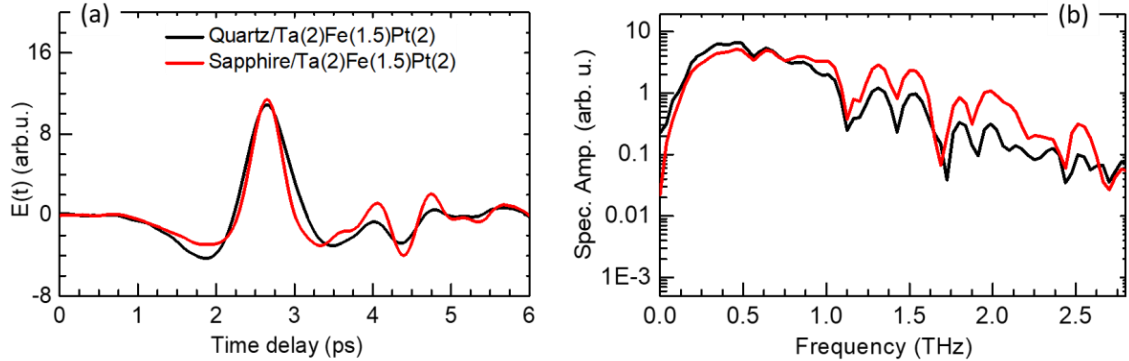

Fig.7. (a) Time-domain THz signal, E(t) and (b) Fourier transformed field spectra of the corresponding THz signals generated from the same tri-layer heterostructure deposited on two different substrates of quartz and sapphire.

## 5. Conclusions

To summarize, we have presented a brief review on the current state and understanding of the spintronic THz emitters. These emitters are suitable combinations of ferromagnetic and nonmagnetic thin films on appropriate substrates, which, upon an optical excitation by visible or NIR femtosecond laser pulses can produce high power and gapless broadband THz radiation. A number of material parameters and the conditions with the optical excitation need to be optimized in the experiments to obtain the maximum possible optical to THz conversion efficiency from the STEs. We also have presented experimental results on the Fe/Pt, Fe/Ta and Ta/Fe/Pt heterostructures to compare the THz generation efficiency from them when excited by sub-100 fs amplified laser pulses centered at 800 nm and 1 kHz repetition rate. It has been shown that Ta(2nm)/Fe(2nm)/Pt(2nm)) tri-layer produces ~40% stronger THz signal than that from the Fe(2nm)/Pt(2nm) bi-layer heterostructure. The same is more than 250% as compared to the Fe/Ta bi-layer structure. The two NM layers used in the tri-layer combination have to have opposite spin-Hall angles. This allows an additional spin to charge current conversion in the tri-layer to improve its THz generation efficiency as compared to the bi-layer structure. Further investigations by improving upon the thickness combinations, spintronic parameters, excitation conditions, appropriate substrates and detection are desirable to fully explore the richness of such new generation THz sources.

**Acknowledgement**

SK acknowledges SERB (DST) and Joint Advanced Technology Center at IIT Delhi for financial support in the research. One of the authors (Sandeep Kumar) acknowledges University Grants Commission, Government of India for Senior Research Fellowship.

**Additional Information**

*Conflict of interest:* All the authors declare no conflict of interest.